\documentclass[runningheads]{llncs}

\usepackage{graphicx} 
\usepackage{pgfplots}
\usepackage{graphics}
\usepackage{array}
\usepackage{tikz}
\usetikzlibrary{positioning}
\usepackage{pgfplotstable}
\pgfplotsset{compat=1.7}
\usepackage{subcaption}
\usepackage{makecell}
\usepackage{caption}
\usepackage{pgf-pie} 
\usepackage{multirow}
\usepackage{multicol}
\usepackage{tabularx}
\usepackage{hyperref}
\usepackage{wrapfig}
\usepackage{comment}
\usepackage{xurl}

\usepackage{amsmath}

\usepackage{listings}
\usepackage[most]{tcolorbox}

\newcommand{\pragformer}[0]{\textsc{PragFormer}}
\newcommand{\ompify}[0]{\textsc{OMPify}}
\newcommand{\dataset}[0]{\textit{Open-OMP-Plus}}

\definecolor{DarkArchtBlue}{HTML}{4b8ac4}
\definecolor{DarkArchtGreen}{HTML}{247d48}
\definecolor{DarkArchtRed}{HTML}{c96565}
\definecolor{DarkArchtPurple}{HTML}{8872c2}
\definecolor{DarkArchtOrange}{HTML}{cf9357}
\definecolor{DarkArchtGrey}{HTML}{878383}
\definecolor{archtBlue}{HTML}{9fc5e8}
\definecolor{archtRed}{HTML}{ea9999}

\usepackage{todonotes}

\usepackage{lipsum}

\usepackage{etoolbox}

\AfterEndEnvironment{figure}{\vskip-1ex}

\begin{document}

\title{Advising OpenMP Parallelization via a Graph-Based Approach with Transformers\vspace{-0.2cm}}
\titlerunning{Advising OpenMP Parallelization with Transformers}
\authorrunning{T. Kadosh, N. Schneider, N. Hasabnis, T. Mattson, Y. Pinter and G. Oren}

\author{Tal Kadosh\inst{1,2} \and
Nadav Schneider\inst{2,3} \and
Niranjan Hasabnis\inst{4} \and
Timothy Mattson\inst{4} \and
Yuval Pinter\inst{1} \and
Gal Oren\inst{5,6}}

\institute{Computer Science Dept, Ben-Gurion University of the Negev, Israel \and Israel Atomic Energy Commission, Tel-Aviv, Israel \and Electrical \& Computer Eng' Dept, Ben-Gurion University of the Negev, Israel \and Intel Labs \and Scientific Computing Center, Nuclear Research Center -- Negev, Israel \and Computer Science Dept, Technion -- Israel Institute of Technology \\ 
\email{talkad@post.bgu.ac.il, nadavsch@post.bgu.ac.il, niranjan.hasabnis@intel.com, timothy.g.mattson@intel.com, uvp@cs.bgu.ac.il, galoren@cs.technion.ac.il}}

\maketitle

\begin{abstract}
\vspace{-0.7cm}
There is an ever-present need for shared memory parallelization schemes to exploit the full potential of multi-core architectures. The most common parallelization API addressing this need today is OpenMP. Nevertheless, writing parallel code manually is complex and effort-intensive. Thus, many deterministic source-to-source (S2S) compilers have emerged, intending to automate the process of translating serial to parallel code. However, recent studies have shown that these compilers are impractical in many scenarios. 
In this work, we combine the latest advancements in the field of AI and natural language processing (NLP) with the vast amount of open-source code to address the problem of automatic parallelization. Specifically, we propose a novel approach, called \ompify{}, to detect and predict the OpenMP pragmas and shared-memory attributes in parallel code, given its serial version. \ompify{} is based on a Transformer-based model that leverages a graph-based representation of source code that exploits the inherent structure of code.
We evaluated our tool by predicting the parallelization pragmas and attributes of a large corpus of (over 54,000) snippets of serial code written in C and C++ languages (\dataset{}).
Our results demonstrate that \ompify{} outperforms existing approaches --- the general-purposed and popular ChatGPT and targeted \pragformer{} models --- in terms of F1 score and accuracy. Specifically, \ompify{} achieves up to 90\% accuracy on commonly-used OpenMP benchmark tests such as NAS, SPEC, and PolyBench. Additionally, we performed an ablation study to assess the impact of different model components and present interesting insights derived from the study. Lastly, we also explored the potential of using data augmentation and curriculum learning techniques to improve the model's robustness and generalization capabilities. 
The dataset and source code necessary for reproducing our results are available at \textcolor{blue}{\url{https://github.com/Scientific-Computing-Lab-NRCN/OMPify}}.\vspace{-0.2cm}

\keywords{NLP \and Code Completion \and OpenMP \and Shared Memory Parallelism \and Transformers\and S2S Compilers \and Code Representations}
\end{abstract}

\section{Introduction}

There is an ever-growing need to develop parallel applications these days.
The ever-growing demand for computing power is leading to various types of complex architectures, including shared-memory multi-core architectures. 
A part of the demand arises from the recent HPCaaS paradigm that has become widespread and available to a broader community of developers~\cite{hpcaastrend}. The services offered as HPCaaS usually depend on the CPU core count and the duration of compute usage. Furthermore, the number of cores per CPU node has increased over the years --- for example, from dozens of physical cores available in GCP's C2 family~\cite{c2family} to hundreds of physical cores available in GCP's future C3 family~\cite{c3family}.


Despite the growing need to write parallel programs, introducing shared-memory parallelization into code remains challenging due to numerous pitfalls. Besides the fact that parallelizing serial code requires extensive knowledge of the code structure and semantics, it also requires the programmer to avoid parallelization pitfalls, such as the need to synchronize simultaneous reads and writes to the same variables (leading to race conditions), as well as making sure that the workload is distributed evenly across the threads and across the system resources (load balancing). In addition, it also requires a high degree of human expertise to comprehend fine details and abstract correlations between variables and different code segments~\cite{datadependency}. It is then unsurprising that the number of parallel programming experts is relatively tiny compared to the growing community of users who can benefit from parallel programs.

The complexity of writing parallel programs is partly addressed by source-to-source (S2S) compilers~\cite{creusillet2009par4all,dave2009cetus,dever2015autopar}, which are compilers that translate code from one programming language to another while preserving the code semantics. These compilers analyze the code for data dependencies that could prevent parallelization and automatically insert appropriate parallelization APIs (such as OpenMP \textit{pragmas}) into it. Nevertheless, these compilers have several major drawbacks~\cite{harel2020source,prema2017identifying,prema2019study}, such as long execution times and limited robustness to the input, even when optimized on runtime~\cite{mosseri2020compar}. More importantly, these compilers require manual development and maintenance efforts, for instance, to support a new programming language or a new specification of parallel programming APIs.


We observed that the recent advances and successes of deep-learning-based Natural Language Processing (NLP) models, such as Transformer architecture and attention mechanism~\cite{vaswani2017attention}, masked language modeling (MLM)~\cite{devlin2018bert}, could help in addressing the limitations of S2S compilers. These models are commonly called large language models (LLMs) because they capture the characteristics of languages. There are already examples of applying these LLMs to programming-related tasks. For example, Codex (based on GPT)~\cite{codex}, a state-of-the-art model that powers GitHub's CoPilot~\cite{githubcopilot}, has shown an interesting application for generating code from natural language prompts. Another example is Google's ML-enhanced code completion tool~\cite{googleadvisor} that can predict possible completions of incomplete code fragments. An internal study conducted by Google has shown the promising potential of these technologies in reducing programming efforts. These technologies are commonly deployed in programming editors and integrated development environments (IDEs) to provide immediate feedback to developers.

Although AI-based programmer assistance tools already exist, to our knowledge, PragFormer~\cite{harel2023learning} is the only AI-based programmer assistance tool that can advise programmers in parallel programming. Specifically, PragFormer uses a Transformer-based architecture to predict if a given serial code could be parallelized (using OpenMP \textit{pragma}) and if private or reduction clause could be applied to it. Specifically, it formulates this problem as multiple binary classification problems, where one problem tackles the need of determining if OpenMP \textit{pragma} could be applied, while the other two tackle the need of determining the need of private and reduction clauses respectively.


While PragFormer has definitely shown interesting perspective towards automated parallel programming, in our experiments with PragFormer, we identified several of its limitations. One of its key limitations is the problem formulation; conceptually, if a serial code cannot be parallelized, then there is no need of determining private/reduction clause. As such, we found that these three are not independent problems, and rather formulating the problem as a multi-label classification problem seems much more intuitive. We address this and a few other limitations in PragFormer to propose a new model, named \ompify{}, that improves upon PragFormer on several fronts. Our experimental evaluation on a corpus of 54,000 \textit{for-loops} mined from GitHub revealed that \ompify{} outperforms PragFormer and several state-of-the-art AI models for code in assisting programmers in parallel programming.


The rest of this article is organized as follows. Section 2 describes related work and provides the necessary background of our work. Section 3 presents the research objectives. Section 4 describes \ompify{} and illustrates our proposed method. Section 5 evaluates our method against previous methods. Finally, section 6 concludes this article and suggests possible extensions of this work.

\section{Related Work}
Initially, the approaches for translating serial code into parallel heavily relied on heuristics and rule-based methods, which often had limited capabilities and robustness~(\S\ref{par:heuristic}). However, with the rapid advancement of deep learning techniques in the field of NLP, along with the easy availability of open-source code, there have been some approaches to apply deep learning techniques to source code~(\S\ref{par:unimodal}). These approaches, however, process source code as text (similar to NLP) and fail to fully exploit the potential of other code representations~(\S\ref{par:multimodal}). By incorporating multiple code representations that capture different aspects of source code, multimodal learning techniques can overcome the limitations of these approaches.

\subsection{Rule-based Methods}\label{par:heuristic}
Several S2S compilers, including Cetus~\cite{dave2009cetus} and Par4All~\cite{creusillet2009par4all}, have emerged in the last decade or so to insert OpenMP pragmas into code automatically. These tools rely on program analysis-based techniques to analyze and identify potential constraints (e.g., loop-carried dependencies) that may restrict the code from being parallelized. The general workflow of S2S compilers can be summarized as follows:
\begin{enumerate}
 \item Create an abstract syntax tree (AST)~\cite{neamtiu2005understanding}, which is a tree representation of the code’s syntactic structure. ASTs are constructed using source code parsers, such as \textit{ANother Tool for Language Recognition} (ANTLR)~\cite{parr2013definitive} or \textit{pycparser}~\cite{bendersky2010pycparser}, etc. 
 \item Apply data dependence algorithms~\cite{fagin1984theory} to ASTs. 
 \item Produce appropriate OpenMP directives based on the data dependence graph.
\end{enumerate}

There are multiple drawbacks associated with the approach of generating ASTs and applying data dependence algorithms. Firstly, creating an AST with a parser can be a challenging task with limited robustness to input due to each programming language's unique syntactic structures that have evolved over the years. Thus, many S2S compilers cannot handle the diverse syntax of programming languages. Moreover, not all parsers are publicly available. As a result, some S2S compilers may fail to produce an AST and analyze the input code. Secondly, data-dependence algorithms can be time-consuming, particularly for large-scale code, since these algorithms are strongly dependent on the size of the AST, which in turn is influenced by the length of the code. Additionally, studies by Harel et al.~\cite{harel2020source} and Prema et al.~\cite{prema2017identifying} have shown that S2S compilers may produce sub-optimal results and even degrade program performance in some cases.

\subsection{Unimodal Machine-Learning Driven Methods}\label{par:unimodal}
Rule-based methods also suffer from another important limitation -- tools relying on these methods require manual programming efforts to add new rules to maintain and update them. However, recent AI-based programming assistance tools have demonstrated that it is possible to reduce manual effort by instead learning the rules from data. Specifically, with the powerful computing devices and vast availability of open-source code as data, these AI-based tools can learn programming rules such as syntax, typing rules~\cite{hasabnis2021controlflag}, etc. Continuing this trend, in recent years, several Transformer-based models have been proposed for various programming-related tasks~\cite{guo2020graphcodebert,niu2022spt}. Typically, these models are pre-trained on massive code corpora containing multiple programming languages (PLs) and then applied to various programming problems~\cite{niu2023empirical}, such as program completion, code search, bug finding, etc., as downstream tasks. One of the common pre-training tasks is masked language modeling (MLM)~\cite{devlin2018bert}.

Previous work by Harel et al.~\cite{harel2023learning} showed the possibility of applying Attention-based models (Transformer) to determine if code can be parallelized with OpenMP. In their work, they introduced \pragformer{}, which is a transformer model based on DeepSCC~\cite{DBLP:deepscc}, which itself is a \textit{RoBERTa} model fine-tuned on a corpus of 225k code snippets written in 21 programming languages (such as Java, Python, C, and C++) collected from Stack Overflow. 

In the parlance of AI-based models, PragFormer formulates the parallel programming assistance problem as \emph{Code Language Processing for Parallelization (CLPP)} task. Specifically, it breaks this task down into three sub-problems: given a serial code (\textit{for-loop}), determine (1) if it can be parallelized (using OpenMP \textit{pragma}), (2) if private clause would be applicable to OpenMP \textit{pragma}, and (3) if reduction class would be applicable to OpenMP \textit{pragma}. It then approaches these three sub-problems independently and formulates them as three separate binary classification problems as below:

\begin{enumerate}
 \item \emph{pragma classification}: Classifying the need for OpenMP \textit{parallel for} \textit{pragma}.
 \item \emph{private clause classification}: Classifying the need for a \textit{private} clause (specifying a variable to be private to each thread in a parallel region). 
 \item \emph{reduction clause classification}: Classifying the need for a \textit{reduction} clause (specifying an operator and a variable to reduce across all threads in a parallel region).
\end{enumerate}

Although PragFormer shows great potential in using Transformer architecture to solve the shared-memory parallelization task, it still suffers from several deficiencies. Primarily, \pragformer{} is based on \textit{RoBERTa}, which is essentially a model for Natural Language (NL) understanding. Applying an NL model to a code-related task is sub-optimal compared to models pre-trained directly on code~\cite{niu2023empirical}. Additionally, \pragformer{} regards source code as a sequence of tokens, ignoring the inherent structure of the code. Intuitively, structural information of code, such as variable dependence information, etc., should provide crucial code semantic information that could improve the code understanding process. Furthermore, the approach of separating the classifications is unintuitive since the tasks of predicting the need for OpenMP \textit{pragma}s and data-sharing attribute clauses are highly correlated --- there will not be a private or reduction clause if there is no need for OpenMP \textit{pragma} at all.

\subsection{Multimodal Machine-Learning Driven Methods}\label{par:multimodal}
While the unimodal ML methods accept source code in only one representation (most commonly as a sequence of tokens), multimodal ML methods realize that other code representations may offer richer semantic information that could improve the accuracy of the models on programming-related tasks. Specifically, multimodel ML models also accept source code in other representations such as AST, control-flow graph (CFG), data-flow graph (DFG), etc. Consequently, many pre-trained machine learning models have been developed, with each model incorporating different code formats into the training process.

Feng et al. presented \textsc{CodeBERT}~\cite{feng2020codebert}, a bimodal Transformer model trained on programming languages alongside natural languages. In their experiments, they used \textit{CodeSearchNet}~\cite{husain2019codesearchnet} dataset, which includes 6.4M code snippets from 6 programming languages (Python, Java, JavaScript, Go, Ruby, and PHP). They compared \textsc{CodeBERT} trained on samples from natural languages and programming languages, \textsc{CodeBERT} trained only on natural languages, and \textsc{RoBERTa}, and showed the superiority of \textsc{CodeBERT} trained on both natural languages and programming languages on several programming-related tasks. Guo et al. created \textsc{GraphCodeBERT}~\cite{guo2020graphcodebert}, a multimodal Transformer model trained on \textit{CodeSearchNet} dataset and input programs as natural language text alongside programming languages and DFG. They showed that DFG enhances the code understanding process compared to \textsc{CodeBERT}. Another multimodal model that exploits the structural aspect of the code is \textsc{SPT-Code}~\cite{niu2022spt} that was proposed by Niu et al. They presented \textsc{SPT-Code}, which is trained on natural language, programming language, and AST, from the \textit{CodeSearchNet} dataset. While \textsc{CodeBERT} and \textsc{GraphCodeBERT} are models that use Transformer encoders, \textsc{SPT-Code} is uses Transformer encoder-decoder architecture. Experiments have shown the superiority of \textsc{SPT-Code} in code generation tasks.

\begin{figure}[!tbp]
\centering
      \includegraphics[width=\textwidth]{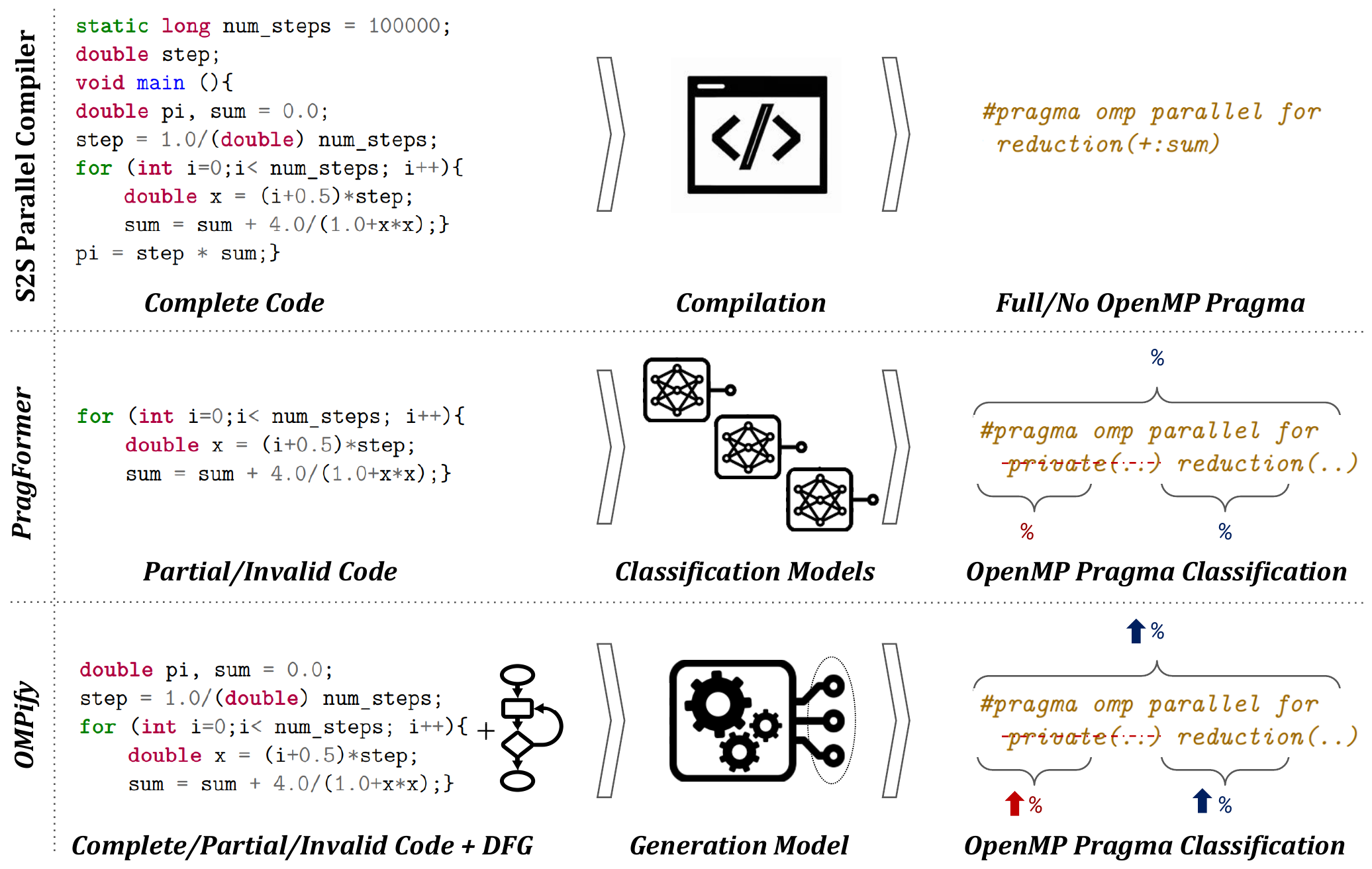}
    \caption{Differences between S2S compilers, \pragformer{} and \ompify{}.}
    \label{fig:diff}
\end{figure}

Drawing inspiration from some of the design choices of multimodal models, we have designed the \ompify{} model also as a multimodal model. Figure~\ref{fig:diff} summarizes the difference between \ompify{} and related works.


\section{Research Objectives}
This paper draws inspiration from PragFormer and approaches the parallel programming assistance problem as a \emph{Code Language Processing for Parallelization (CLPP)} task. Nevertheless, we improve upon PragFormer by posing the following research questions that are designed to evaluate the limitations of PragFormer discussed earlier in Related Work (\S\ref{par:unimodal}).



\vspace{-1ex}
\subsection*{RQ1: \normalfont{\textit{Which code representations impact the CLPP task?}}}\label{sec:rq1}
\vspace{-1ex}

Given the discussion of different code representations, this question focuses on assessing the influence of various code modalities on code comprehension, particularly in CLPP tasks. We will evaluate the effectiveness of the previously mentioned multimodal models on our dataset.

\vspace{-1ex}
\subsection*{RQ2: \normalfont{\textit{Does the scope of the for-loop from input serial code matter for performance on CLPP task?}}}\label{sec:rq2}
\vspace{-1ex}

Conceptually, the semantics of a \textit{for-loop} from serial code heavily relies on its \emph{context}. This question assesses the impact of different context lengths on the performance of various multimodal models on CLPP tasks.

\vspace{-1ex}
\subsection*{RQ3: \normalfont{\textit{Can code augmentation improve model's performance on CLPP task?}}}\label{sec:rq3}
\vspace{-1ex}

We will investigate the potential of code augmentation techniques, specifically variable name replacement, to improve the performance of existing models on CLPP tasks.

\vspace{-1ex}
\subsection*{RQ4: \normalfont{\textit{Will multi-label classification based formulation for solving CLPP task perform better than PragFormer's multiple binary-classification based formulation?}}}\label{sec:rq4}
\vspace{-1ex}

While \pragformer{} employed three binary classification-based models to predict the requirement for an OpenMP \textit{pragma} and whether it should include a work-sharing construct, we hypothesize that these predictions are interdependent and potentially benefit each other. We evaluate this hypothesis by formulating a multi-label classification problem and developing a single generative model for the CLPP task. We then compare our model against PragFormer and other state-of-the-art multimodel models.

\section{OMPify}
This section describes the model architecture, its input, the code representations, and the fine-tuning process. \ompify{} predicts the need for both OpenMP \textit{pragma} and shared-memory attributes (\textit{private} and \textit{reduction}) simultaneously, allowing the model to learn inter-dependencies between these tasks.

\subsection{Model}
\begin{figure}
\centering
      \includegraphics[width=\textwidth]{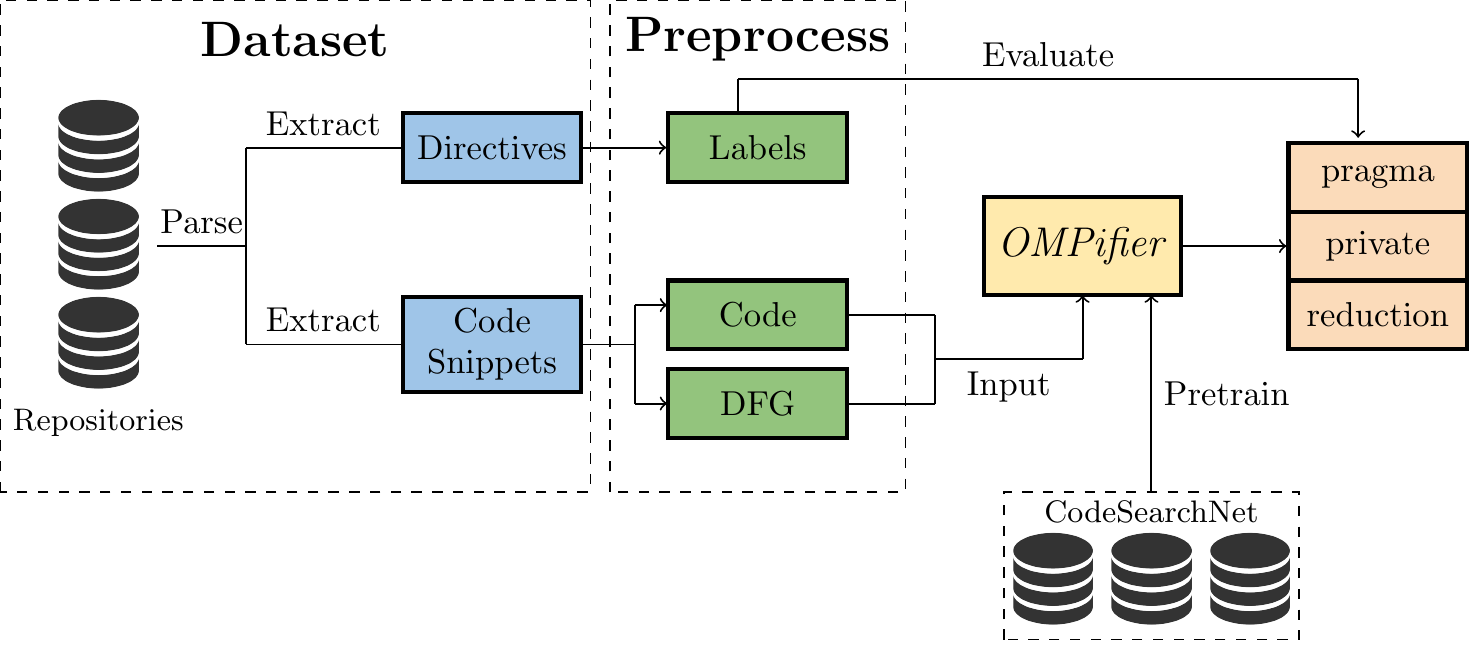}
    \caption{Overview of \ompify{} training process.}
    \label{fig:teaser}
\end{figure}

\ompify{}~(\autoref{fig:teaser}) is a Transformer-based, multimodal model that utilizes a graph-based representation of source code.
\ompify{} is based on \textsc{GraphCodeBERT}~\cite{guo2020graphcodebert}, a pre-trained model for programming languages that considers the inherent structure of the code by accepting source code along with its DFG. \ompify{} is composed of \textsc{GraphCodeBERT} and a fully connected layer. This architecture allows \ompify{} to perform multi-label classification, where each task is individually classified.

\subsection{Model Input}
The model's inputs are two code modalities: the actual source code as a sequence of tokens and the serialized DFG.

\begin{figure}[ht!]
\centering
      \includegraphics[width=\textwidth]{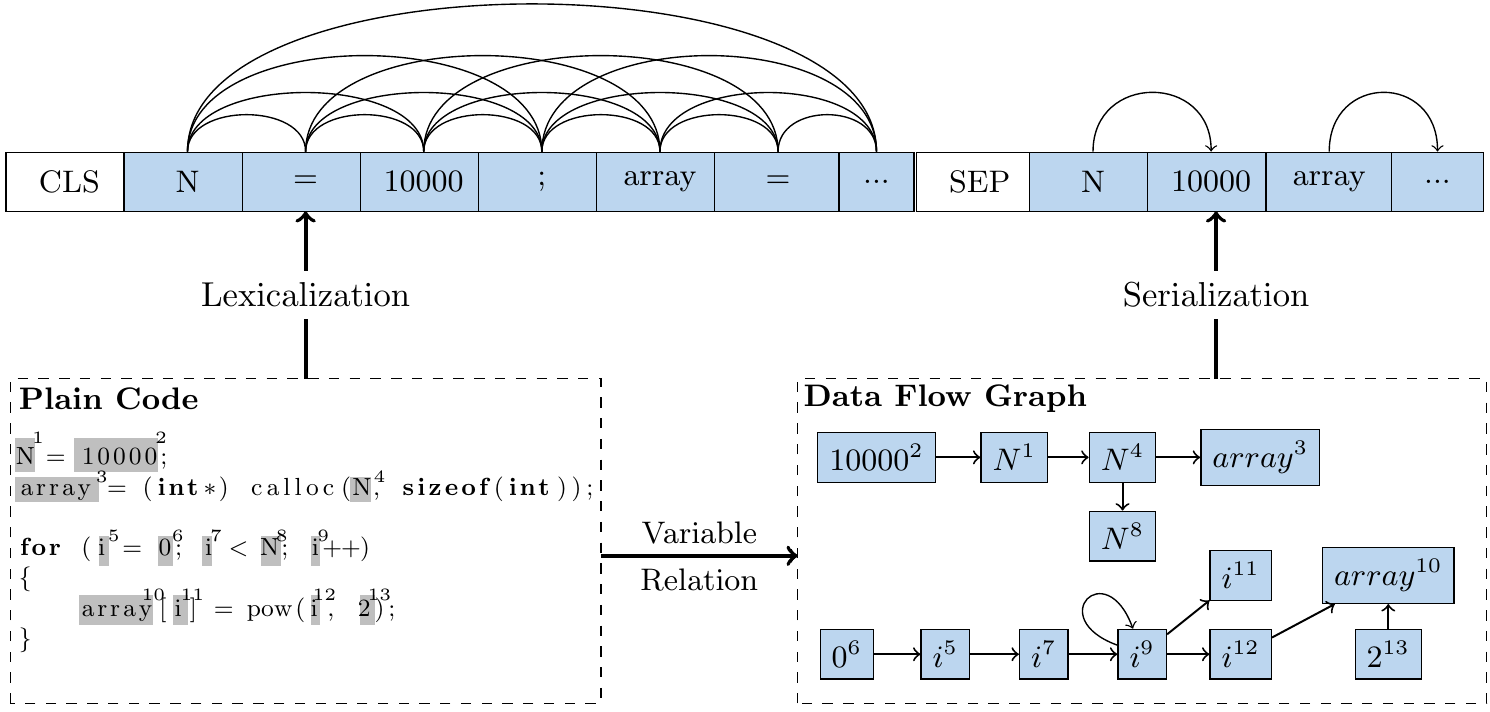}
    \caption{The input format for a C code snippet.}
    \vspace{-0.5cm}
    \label{fig:input_format}
\end{figure}

\begin{itemize}
    \item \noindent\textbf{Code Tokens.} As shown in \autoref{fig:input_format}, the first part of the input to \ompify{} consists of a sequence of code tokens. Feeding source code as a sequence of strings to Transformer-based models does not work well in practice. A common approach, in this case, is to train a tokenizer that maps strings of tokens into unique IDs (numbers), such that strings that are semantically closer have their IDs closer also. We used the tokenizer provided by \textsc{GraphCodeBERT} to generate a sequence of token IDs.
    \item \noindent\textbf{Serialized DFG.} The second part of the input, which is the serialized DFG, is created by converting the code into an AST using TreeSitter parser~\footnote{\textcolor{blue}{\url{https://github.com/tree-sitter/tree-sitter}}}. We extract variables and their data dependence relationships from an AST to generate a DFG. The DFG nodes are serialized in the program ordered and serve as the model input.
    \item \noindent\textbf{Attention Mask.} \autoref{fig:input_format} provides an overview of the token connections, showcasing the interconnections between the tokens. Whereas the code tokens attend to each other during the self-attention mechanism~\cite{DBLP:journals/corr/VaswaniSPUJGKP17}, when dealing with DFG, we aim to disregard attention between variables that are not connected. To achieve this, we employed the masked attention approach described by Guo et al.~\cite{guo2020graphcodebert}.
    During the computation of self-attention, we utilized an attention mask, denoted as $M$, which contains zeros in positions $(i, j)$ where we want the tokens to attend to each other, and contains $-inf$ to prevent certain relations. The self-attention computation can be represented as follows: 

    \vspace{-0.7cm}
    
    \begin{equation}\label{eq:sa}
        \resizebox{0.5\linewidth}{!}{$SelfAttention = \text{softmax}(QK/d + M)V$}
    \end{equation}

\end{itemize}

\vspace{-0.4cm}
\subsection{Fine Tuning}

Many studies have demonstrated the advantages of implementing data augmentation techniques to enhance the performance of deep learning models~\cite{rebuffi2021data,wang2022bridging}. Data augmentation techniques are typically applied to the training set to increase the diversity of input. Commonly-used augmentation techniques include \textit{variable renaming}, \textit{dead store}, and \textit{constant replacement}.

Despite the effectiveness of data augmentations, several studies~\cite{henke2022semantic,quiring2019misleading} showed that deep learning models are vulnerable to adversarial examples, i.e., minor changes to code can result in significant performance degradation. To address this issue, as many studies have suggested~\cite{guo2018curriculumnet,platanios2019competence,wang2022bridging} leveraging a curriculum learning (CL) technique that involves the gradual introduction of data augmentation techniques. 
Specifically, we applied \textit{variable renaming} as a data augmentation technique.
We followed a gradual approach, starting with the original data without any augmentation during the first epoch. In each subsequent epoch, we augmented the original data by progressively increasing the proportion of renamed variables. Specifically, during the second epoch, we renamed 10\% of the variables for each sample in the original training set. In the third epoch, we continued this approach and renamed 20\% of the variables per sample. As we progressed to the fourth epoch, we further increased the update ratio, renaming 30\% of the variables per sample. Finally, starting from the fifth epoch and throughout the remaining epochs, we consistently maintained an update ratio of 40\%, resulting in the renaming of 40\% of the variables per sample in each subsequent epoch.



\section{Experimental Results}
To evaluate the effectiveness of our proposed model, \ompify{}, we conducted several experiments to answer our research questions. All experiments were conducted on an NVIDIA A100 GPU. Furthermore, for the sake of consistency, we utilized the original implementations of the models as presented in their respective papers.

\subsection{Dataset \& Preprocessing}
In our work, we developed a novel dataset, named \dataset{}, comprising more than 54,000 code snippets from C and C++ for OpenMP analysis (\autoref{fig:dataset}). The dataset was collected from \url{github.com} using the \textit{github-clone-all}\footnote{\textcolor{blue}{\url{http://github.com/rhysd/github-clone-all}}} script, which enables searching for repositories that satisfy specific criteria. We used this tool to locate all repositories that include C or C++ files and also feature the term ``OpenMP'' in their title, description, or README.

\begin{table}[!htbp]
    \centering
    \begin{subfigure}{0.25\textwidth}
        \centering
        \begin{footnotesize}
        \begin{tabular}{|c|c|c|}
            \hline
            \textbf{Description} & \textbf{C} & \textbf{C++}\\
            \hline
            With OpenMP & 14,906 & 8,241 \\ \hline
            Without OpenMP & 17,193 & 14,323 \\ \hline
            Total & 32,099 & 22,564 \\
            \hline
        \end{tabular}
        \end{footnotesize}
        \captionsetup{width=1.5\textwidth}
        \caption{Number of loops paralleled with OpenMP for each programming language (C and C++).}

        \label{tab:samples_dist}
    \end{subfigure}%
    \hfill \hspace{10ex}
    \begin{subfigure}{0.25\textwidth}
        \centering
        \begin{footnotesize}
        \begin{tabular}{|c|c|}
        \hline
        \textbf{Clauses} & \textbf{Amount}\\
        \hline
        \textit{private} & 6,758  \\ \hline
        \textit{reduction} & 3,267 \\ \hline
        Total & 10,025 \\
        \hline
        \end{tabular}
        \end{footnotesize}
        \caption{Number of common OpenMP shared memory attributes.}
        \label{tab:clauses}
    \end{subfigure}\hfill
    \begin{subfigure}{0.25\textwidth}
        \centering
        \begin{footnotesize}
        \begin{tabular}{|c|c|}
        \hline
        \textbf{\# Lines} & \textbf{Amount}\\
        \hline
        \textit{$<$ 15} & 40,745  \\ \hline
        \textit{16-50} & 10,607 \\ \hline
        \textit{$>$ 50} & 3,311 \\
        \hline
        \end{tabular}
        \end{footnotesize}
        \caption{Code snippet length in \dataset{}.}
        \label{tab:clauses}
    \end{subfigure}\hfill
    \caption{The distribution of each class for each programming language.}
    \vspace{-0.5cm}
    \label{fig:dataset}  
\end{table}

To minimize the noise in our dataset, we employed inclusion and exclusion criteria inspired by Harel et al.~\cite{harel2023learning}. Specifically, we included only C/C++ files that contained OpenMP \textit{pragma}s in their code. This criteria operates on the assumption that the developers were aware of OpenMP parallelization and that any non-parallelized loops intentionally have not been parallelized. We excluded duplicates, empty loops, and loops that used \textit{barrier}, \textit{critical}, or \textit{atomic pragma}s, which can be bottlenecks on code execution and are not optimal samples. 

Once we identified files that contained OpenMP pragmas, we parsed them using \emph{pycparser}~\cite{bendersky2010pycparser} parser, which converts the code into an AST format. Each sample in the dataset comprises several fields, such as the plain \textit{for-loop} code, its corresponding \textit{pragma} (if any), the AST of the \textit{for-loop}, the AST of the functions called within the \textit{for-loop}, the declaration of each variable used in the \textit{for-loop}, all the assignment instructions from the context of the loop that involves each of the variables used in the loop, and the DFG of the \textit{for-loop} and its extended scope. By analyzing the AST, we can extract code structures such as loops and identify the relevant functions and variables from its outer scope.


To evaluate the performance of our model, we divided the dataset into three sets: train, validation, and test, using standard 80-10-10 split. Additionally, we collected three benchmarks that were known to use OpenMP correctly, namely NAS~\cite{bailey1991parallel}, PolyBench~\cite{polybench}, and SPEC~\cite{spec_omp}, and used them to further test our model. To avoid fair evaluation, we removed from the training set any samples that could be found in the benchmarks.

%
%

\subsection{Results}

We now present the results of our experiments to answer the research questions.


\subsubsection{RQ1: Code modalities.}
\label{RQ1res}

\begin{wraptable}{r}{7cm}
\centering
    \vspace{-0.7cm}

    \begin{footnotesize}
    \begin{tabular}{|c|c|c|c|}
    \hline
    \multicolumn{1}{|c|}{\multirow{2}{*}{\textbf{Model Name}}} & \multicolumn{3}{c|}{\textbf{Metrics}} \\ \cline{2-4} 
    \multicolumn{1}{|c|}{} & \textbf{P} & \textbf{R} & \textbf{Acc} \\ \hline

        PragFormer & 0.826 & 0.780 & 0.830 \\  \hline
        CodeBERT & \textbf{0.848} & 0.813 & 0.852 \\  \hline
        SPT-Code & 0.812 & 0.784 & 0.831 \\  \hline
        SPT-Code (code only) & 0.792 & 0.786 & 0.820 \\  \hline
        GraphCodeBERT& 0.836 & \textbf{0.835} & \textbf{0.862} \\  \hline
        GraphCodeBERT (code only) & 0.834 & 0.833 & 0.861 \\  \hline
    \end{tabular}
    \end{footnotesize}

    \caption{Effect of different code modalities on the task of \textit{pragma classification}. (P=Precision, R=Recall, Acc=Accuracy)}
    \vspace{-0.7cm}
    \label{tab:comparison}
\end{wraptable}





To compare the various code modalities, we utilized three distinct models. \textsc{CodeBERT} was pre-trained on natural language (NL) and programming language (PL), \textsc{SPT-Code} was pre-trained on PL and AST, and \textsc{GraphCodeBERT} was pre-trained on PL and DFG. To apply these models to \textit{pragma classification} task, we added a fully connected layer of size two and a SoftMax layer at the end of these models and fine-tuned them using our corpus, \dataset{}. We also included \pragformer{}, which was fine-tuned on our corpus, for comparison with these models.

Based on the results presented in \autoref{tab:comparison}, it can be inferred that the use of multimodal models, which combine code representations such as AST or DFG with the original code, has a positive impact on the performance of the model in the \textit{pragma classification} task. However, despite being trained on the same dataset, i.e., \textit{CodeSearchNet}, \textsc{CodeBERT}, \textsc{SPT-Code}, and \textsc{GraphCodeBERT} achieved significantly different performance, with \textsc{GraphCodeBERT} outperforming others. This indicates that the DFG representation and pretraining tasks proposed by Guo et al.~\cite{guo2020graphcodebert} were more beneficial this task. It is worth noting that while the use of AST in \textsc{SPT-Code} significantly improves performance compared to using only the code, the DFG in \textsc{GraphCodeBERT} has only a minimal impact on performance. This could be due to the DFG representation's inability to effectively capture the relationship between arrays and their indexing. As shown in \autoref{fig:input_format}, there is no direct connection between the array and the index variable $i^{11}$. In scientific codes, the relationship between the array and the index is often critical as it determines the feasibility of parallelization.

\subsubsection{RQ2: Extended scope.}
\label{RQ2res}


    
        
    

In this experiment, we aimed to investigate the impact of extended scope on the performance of \ompify{} in solving the CLPP task. To achieve this, we trained \ompify{} on two distinct corpora: one comprising only the \textit{for-loop} structured block, and the other consisting of the extended version that includes the surrounding scope of the \textit{for-loop}, incorporating assignments to variables used within the loop~(\autoref{tab:length_comparison}).
The results, as presented in Table \ref{tab:extended_scope}, demonstrate the effect of including the outside scope in determining the necessity of OpenMP \textit{pragma}. The observed increase in recall indicates that the model exhibits improved identification of \textit{for-loop}s requiring OpenMP \textit{pragma}, resulting in fewer false negatives. This finding suggests that considering the outside scope provides valuable information for accurately identifying the need for \textit{pragma} in \textit{for-loop}s.

\vspace{-0.8cm}
\begin{figure}[!h]
    \centering
    \begin{minipage}[b]{0.4\textwidth}
        \centering
        \begin{tabular}{|c|c|c|c|}
        \hline
        \textbf{Data Type} & \textbf{P} & \textbf{R} & \textbf{Acc} \\ \hline
        No Scope & \textbf{0.833} & 0.831 & 0.860 \\  \hline
        With Scope & 0.829 & \textbf{0.844} & \textbf{0.863} \\  \hline
        \end{tabular}
        \caption{Effect of \emph{context}. (P=Precision, R=Recall, Acc=Accuracy)}
        \label{tab:extended_scope}
    \end{minipage}%
    \begin{minipage}[b]{0.6\textwidth}
        \centering
        \begin{tikzpicture}
            \begin{axis}[
                ybar,
                width=6.8cm,
                height=3.6cm,
                ymin=0,
                ylabel={\# Lines},
                xtick=data,
                xticklabels={$<10$, $11-50$, $50-100$, $>100$},
                bar width=0.4cm,
                legend style={at={(0.95,0.95)}, anchor=north east},
                ]
            \addplot[draw=black, fill=archtBlue!140] coordinates {(1, 9155) (2, 23607) (3, 14686) (4, 4985)};
            \addplot[draw=black, fill=archtRed!140] coordinates {(1, 31187) (2, 14998) (3, 5153) (4, 1613)};
            \legend{Extended Scope, No Scope}
            \end{axis}
        \end{tikzpicture}
        \caption{Code length comparison.}
        \label{tab:length_comparison}
        
    \end{minipage}
\end{figure}

  
      
          
          
  

\subsubsection{RQ3: Data augmentation.}
\label{RQ3res}

  
      

\begin{wraptable}{r}{7cm}
\centering
    \begin{footnotesize}
    \vspace{-0.7cm}
    \begin{tabular}{|c|c|c|c|c|}
    \hline
    \multicolumn{1}{|c|}{\multirow{2}{*}{\textbf{Model}}} & \multicolumn{1}{c|}{\textbf{Augmen-}} & \multicolumn{3}{c|}{\textbf{Metrics}} \\ \cline{3-5} 
    \multicolumn{1}{|c|}{} & \multicolumn{1}{c|}{\textbf{tation}} & \textbf{P} & \textbf{R} & \textbf{Acc} \\ \Xhline{2.5\arrayrulewidth}
    \multirow{3}{*}{PragFormer} & original & 0.793 & \textbf{0.847} & 0.841 \\  \cline{2-5}
     & curriculum & 0.825 & 0.815 & 0.848 \\ \cline{2-5}
     & replaced & 0.727 & 0.826 & 0.794 \\ \Xhline{2.5\arrayrulewidth}
     \multirow{3}{*}{\makecell{GraphCode-\\ BERT}} & original & \textbf{0.851} & 0.841 & 0.870 \\  \cline{2-5}
     & curriculum & 0.849 & 0.846 & \textbf{0.872} \\  \cline{2-5}
     & replaced & 0.838 & 0.781 & 0.843 \\ \hline
    \end{tabular}
    \end{footnotesize}
    
\caption{Effect of data augmentation techniques. (P=Precision, R=Recall, Acc=Accuracy)}
\vspace{-0.9cm}
\label{tab:data_augmentation}
\end{wraptable}

\vspace{-0.8cm}
Through evaluating the impact of data augmentation techniques on performance, we investigated the effectiveness of \pragformer{} and \textsc{GraphCodeBERT} in the binary classification task of OpenMP \textit{pragma} classification. The results of this experiment are presented in \autoref{tab:data_augmentation}. We employed the \textit{variable renaming} augmentation, where each variable was replaced with \textit{var}, concatenated with a random index number. This augmentation is referred as \textit{replaced} in the table. The results reveal the vulnerability of these models to adversarial examples created by fully replacing variable names, leading to degraded performance compared to the unmodified variables. However, by gradually introducing code augmentations using the curriculum learning method (referred as \textit{curriculum} in the table), we observed improved accuracy.






\vspace{-0.cm}
\subsubsection{RQ4: Multi-label classification.}
\label{RQ4res}

      
      
       
      
  
\begin{wraptable}[8]{r}{7cm}
\centering
    \begin{footnotesize}
    \vspace{-1.4cm}
    \begin{tabular}{|c|c|c|c|c|}
    \hline
    \multicolumn{1}{|c|}{\multirow{2}{*}{\textbf{Model}}} & \multicolumn{1}{c|}{\multirow{2}{*}{\textbf{Task}}} & \multicolumn{3}{c|}{\textbf{Metrics}} \\ \cline{3-5} 
    \multicolumn{1}{|c|}{} & \multicolumn{1}{c|}{} & \textbf{P} & \textbf{R} & \textbf{Acc} \\ \Xhline{2.5\arrayrulewidth}
    \multirow{3}{*}{PragFormer} & \textit{pragma} & 0.793 & 0.847 & 0.841 \\ \cline{2-5}
     & \textit{private} & 0.716 & 0.663 & 0.924 \\  \cline{2-5}
     & \textit{reduction} & 0.632 & 0.598 & 0.953 \\ \cline{2-5} \Xhline{2.5\arrayrulewidth}
    
    \multirow{3}{*}{\makecell{GraphCode-\\ BERT \\ (Separated)}} & \textit{pragma} & \textbf{0.850} & 0.841 & 0.870 \\ \cline{2-5}
     & \textit{private} & \textbf{0.768} & 0.684 & 0.937 \\  \cline{2-5}
     & \textit{reduction} & \textbf{0.690} & 0.688 & 0.963 \\ \cline{2-5} \Xhline{2.5\arrayrulewidth}
     
    \multirow{3}{*}{OMPify} & \textit{pragma} & 0.849 & \textbf{0.848} & \textbf{0.872} \\ \cline{2-5}
     & \textit{private} & 0.755 & \textbf{0.689} & \textbf{0.938} \\ \cline{2-5}
     & \textit{reduction} & \textbf{0.690} & \textbf{0.700} & \textbf{0.966} \\ \hline
    \end{tabular}
    \end{footnotesize}
    
    \caption{Effect of multi-label classification problem formulation. (P=Precision, R=Recall, Acc=Accuracy)}
    \vspace{-0.7cm}
    \label{tab:multiclass_performance}
\end{wraptable}

\autoref{tab:multiclass_performance}  shows the results of \pragformer{}, \textsc{GraphCodeBERT}, and \ompify{} when applied to all the three tasks of \textit{pragma} classification, classification of \textit{private} clause, and \textit{reduction} clause.

\noindent{}Note that \ompify{} approaches all three tasks together as a multi-label classification problem, while \pragformer{}, \textsc{GraphCodeBERT} approach each task independently. In the table, we present the result of \ompify{} for the combined task but split the results according to labels.

The results convey that \ompify{} achieves significantly better performance compared to \textsc{PragFormer}, underscoring the hypothesis that these three tasks are not independent. In addition, our model slightly outperforms the \textsc{GraphCodeBERT} model. The results show a significant improvement in recall for \ompify{} for all three tasks, with a major decrease in the number of false negative predictions. In our context, a false negative prediction means that a sample is incorrectly classified as not requiring \textit{pragma}, \textit{private}, or \textit{reduction}. Therefore, the unified prediction strategy of \ompify{} can better identify samples that require \textit{pragma}, \textit{private}, or \textit{reduction}. This suggests that the understanding of each task contributes to the overall prediction. For instance, if \ompify{} predicts the need for \textit{pragma}, it will also influence the prediction of shared-memory attributes, which may also appear in the \textit{pragma}.


\subsubsection{Real-world benchmarks.}
\label{sec:bench}

\begin{wraptable}{r}{7cm}
    \centering
    \begin{footnotesize}
    \vspace{-0.7cm}
    \begin{tabular}{|c|c|c|c|c|}
    \hline
    \multirow{2}{*}{\makecell{\textbf{Bench-} \\ \textbf{mark}}} & \multirow{2}{*}{\makecell{\textbf{With} \\ \textbf{OMP}}} & \multirow{2}{*}{\makecell{\textbf{Without} \\ \textbf{OMP}}} & \multirow{2}{*}{\makecell{\textbf{\textit{priv-}} \\ \textbf{\textit{ate}}}} &
    \multirow{2}{*}{\makecell{\textbf{\textit{reduc-}} \\ \textbf{\textit{tion}}}} \\
    & & & & \\ \hline
    NAS & 166 & 146 & 12 & 2 \\ \hline
    PolyBench & 63 & 85 & 36 & 0 \\ \hline
    SPEC & 157 & 1,000 & 1 & 0 \\ \hline
    \end{tabular}
    \end{footnotesize}

    \caption{Benchmark statistics}
    \vspace{-0.7cm}
    \label{tab:benchmark_stats}
\end{wraptable}

In order to test the performance of \ompify{} on real-world programs, we obtained C/C++ programs that were using OpenMP \textit{pragma} from three scientific code benchmarks, namely, NAS, SPEC, and PolyBench. \autoref{tab:benchmark_stats} shows the statistics of the collected programs. These benchmarks are manually-written as parallel programs using OpenMP, so they serve as a good test case for \ompify{}. As a comparison, we applied \pragformer{} to the same test. Our model exhibited a significant increase in performance when compared to \pragformer{}~(\autoref{tab:dist}).

\begin{wraptable}{r}{7cm}
    \centering
    \begin{footnotesize}
    \vspace{-0.7cm}
    \begin{tabular}{|c|c|c|c|c|}
        \hline
        \multirow{2}{*}{\makecell{\textbf{Bench-} \\ \textbf{mark}}} &
        \multirow{2}{*}{\textbf{Model}} &
        \multicolumn{3}{c|}{\textbf{Metrics}} \\ \cline{3-5}
        
         & & \textbf{P} & \textbf{R} & \textbf{Acc} \\ \Xhline{2.5\arrayrulewidth}

        \multirow{2}{*}{SPEC} & \pragformer{} & 0.445 & 0.802 & 0.837 \\ \cline{2-5}
            & \ompify{} & \textbf{0.572} & \textbf{0.854} & \textbf{0.894} \\ \Xhline{2.5\arrayrulewidth}

        \multirow{2}{*}{PolyBench} & \pragformer{} & 0.703 & 0.301 & 0.648 \\ \cline{2-5}
            & \ompify{} & \textbf{0.836} & \textbf{0.810} & \textbf{0.851} \\ \Xhline{2.5\arrayrulewidth}

        \multirow{2}{*}{NAS} & \pragformer{} & 0.635 & 0.734 & 0.634 \\ \cline{2-5}
           & \ompify{} & \textbf{0.731} & \textbf{0.886} & \textbf{0.766} \\ \Xhline{2.5\arrayrulewidth}

        \multirow{3}{*}{\makecell{2500 \\ examples}} & ChatGPT & 0.401 & \textbf{0.913} & 0.401 \\ \cline{2-5}
           & \pragformer{} & 0.815 & 0.721 & 0.817 \\ \cline{2-5}
           & \ompify{} & \textbf{0.839} & 0.818 & \textbf{0.860} \\ \hline
    \end{tabular}
    \end{footnotesize}
    \caption{Comparison on different benchmarks. (P=Precision, R=Recall, Acc=Accuracy)}
    \vspace{-0.7cm}
    \label{tab:dist}
\end{wraptable}

Moreover, given the recent popularity of ChatGPT~\cite{chatgpt} in programming-related tasks, we decided to evaluate it on our CLPP task. For this evaluation, we randomly sampled 2500 test inputs from our test dataset. We then fed those test programs to ChatGPT one by one and then used the prompt \textit{``Generate the optimal OpenMP pragma if possible''} to check if ChatGPT's response matches with the expected label for the test program. Although ChatGPT performs well on various NLP tasks, it performed poorly in our specific task, often suggesting the use of OpenMP \textit{pragma} even when it was not applicable.

\section{Conclusions \& Future Work}
This paper aims to investigate the potential of multimodal models in accurately predicting the need for shared-memory parallelization in code. Our research discovered that incorporating additional code representations, such as ASTs and DFGs, significantly improves their performance compared to models that rely solely on the original code.
Building upon this knowledge, we introduced a novel model called \ompify{}, based on \textsc{GraphCodeBERT}. \ompify{} takes advantage of the inter-dependencies between the task of predicting the need for parallelization and the prediction of shared-memory attributes, such as \textit{private} and \textit{reduction} variables. By leveraging these relationships, \ompify{} demonstrates enhanced accuracy and robustness in determining the need for shared-memory parallelization.
In addition to developing the \ompify{} model, we also constructed a comprehensive database called \dataset{}. This database includes the \textit{for-loop} itself and extends its scope to include assignment statements of variables found within the \textit{for-loop}. By incorporating this extended scope, we demonstrate that \ompify{} can effectively utilize this additional information to improve its predictions further.

For future research, we aim to address several areas of improvement. Firstly, since the multimodal models analyzed in \textsc{RQ1} were not pre-trained on C/C++ programming languages, there is a potential for enhancing their performance by pretraining them on datasets that include C/C++ code. This approach can contribute to better code understanding and comprehension.
Additionally, in \textsc{RQ4}, we observed improvements in multi-label prediction. To further enhance this aspect, we intend to explore the conversion of the multi-label prediction problem into \textit{pragma} generation. By generating \textit{pragma}s directly, we can achieve more precise and fine-grained control over parallelization tasks.
Furthermore, an important question arises regarding the correctness of the generated \textit{pragma}s. To address this concern, we plan to investigate techniques and approaches for evaluating the accuracy and correctness of the generated \textit{pragma}s. 

\hfill \\
\noindent
\textbf{Acknowledgments}: 
This research was supported by the Israeli Council for Higher Education (CHE) via the Data Science Research Center, Ben-Gurion University of the Negev, Israel; Intel Corporation (oneAPI CoE program); and the Lynn and William Frankel Center for Computer Science. Computational support was provided by the NegevHPC project~\cite{negevhpc} and Intel Developer Cloud~\cite{intel-cloud}. The authors thank Re'em Harel, Israel Hen, and Gabi Dadush for their help and support.

\newpage

\bibliographystyle{splncs04}
\bibliography{base}

\begin{thebibliography}{10}
\providecommand{\url}[1]{\texttt{#1}}
\providecommand{\urlprefix}{URL }
\providecommand{\doi}[1]{https://doi.org/#1}

\bibitem{datadependency}
{Automatic Parallelism and Data Dependency}.
  \url{https://web.archive.org/web/20140714111836/http://blitzprog.org/posts/automatic-parallelism-and-data-dependency
  }, [Online]

\bibitem{c2family}
{Compute-optimized machine family}.
  \url{https://cloud.google.com/compute/docs/compute-optimized-machines },
  [Online]

\bibitem{githubcopilot}
{GitHub Copilot}. \url{https://github.com/features/copilot }, [Online]

\bibitem{hpcaastrend}
{High Performance Computing as a Service Market Forecast}.
  \url{https://www.alliedmarketresearch.com/high-performance-computing-as-a-service-market
  }, [Online]

\bibitem{googleadvisor}
Ml-enhanced code completion improves developer productivity.
  \url{https://ai.googleblog.com/2022/07/ml-enhanced-code-completion-improves.html
  }, [Online]

\bibitem{negevhpc}
{NegevHPC Project}. \url{https://www.negevhpc.com}, [Online]

\bibitem{polybench}
{PolyBench Benchmarks}.
  \url{https://web.cse.ohio-state.edu/~pouchet.2/software/polybench/}, [Online]

\bibitem{spec_omp}
{SPEC-OMP2012 website}. \url{https://www.spec.org/omp2012/ }, [Online]

\bibitem{c3family}
{The next wave of Google Cloud infrastructure innovation: New C3 VM and
  Hyperdisk}.
  \url{https://cloud.google.com/blog/products/compute/introducing-c3-machines-with-googles-custom-intel-ipu
  }, [Online]

\bibitem{chatgpt}
{ChatGPT}. \url{https://chat.openai.com/} (2023), [Online]

\bibitem{bailey1991parallel}
Bailey, D.H., Barszcz, E., Barton, J.T., Browning, D.S., Carter, R.L., Dagum,
  L., Fatoohi, R.A., Frederickson, P.O., Lasinski, T.A., Schreiber, R.S.,
  et~al.: The nas parallel benchmarks. The International Journal of
  Supercomputing Applications  \textbf{5}(3),  63--73 (1991)

\bibitem{bendersky2010pycparser}
Bendersky, E., et~al.: Pycparser (2010)

\bibitem{codex}
Chen, M., Tworek, J., Jun, H., Yuan, Q., Pinto, H.P.d.O., Kaplan, J., Edwards,
  H., Burda, Y., Joseph, N., Brockman, G., et~al.: Evaluating large language
  models trained on code. arXiv preprint arXiv:2107.03374  (2021)

\bibitem{creusillet2009par4all}
Creusillet, B., Keryell, R., Even, S., Guelton, S., Irigoin, F.: Par4all:
  Auto-parallelizing c and fortran for the cuda architecture  (2009)

\bibitem{dave2009cetus}
Dave, C., Bae, H., Min, S.J., Lee, S., Eigenmann, R., Midkiff, S.: Cetus: A
  source-to-source compiler infrastructure for multicores. Computer
  \textbf{42}(12) (2009)

\bibitem{dever2015autopar}
Dever, M.: AutoPar: automating the parallelization of functional programs.
  Ph.D. thesis, Dublin City University (2015)

\bibitem{devlin2018bert}
Devlin, J., Chang, M.W., Lee, K., Toutanova, K.: {BERT}: Pre-training of deep
  bidirectional transformers for language understanding. In: Proceedings of the
  2019 Conference of the North {A}merican Chapter of the Association for
  Computational Linguistics: Human Language Technologies, Volume 1 (Long and
  Short Papers). pp. 4171--4186. Association for Computational Linguistics,
  Minneapolis, Minnesota (Jun 2019). \doi{10.18653/v1/N19-1423},
  \url{https://aclanthology.org/N19-1423}

\bibitem{fagin1984theory}
Fagin, R., Vardi, M.Y.: The theory of data dependencies: a survey. IBM Thomas
  J. Watson Research Division (1984)

\bibitem{feng2020codebert}
Feng, Z., Guo, D., Tang, D., Duan, N., Feng, X., Gong, M., Shou, L., Qin, B.,
  Liu, T., Jiang, D., et~al.: Codebert: A pre-trained model for programming and
  natural languages. arXiv preprint arXiv:2002.08155  (2020)

\bibitem{guo2020graphcodebert}
Guo, D., Ren, S., Lu, S., Feng, Z., Tang, D., Liu, S., Zhou, L., Duan, N.,
  Svyatkovskiy, A., Fu, S., et~al.: Graphcodebert: Pre-training code
  representations with data flow. arXiv preprint arXiv:2009.08366  (2020)

\bibitem{guo2018curriculumnet}
Guo, S., Huang, W., Zhang, H., Zhuang, C., Dong, D., Scott, M.R., Huang, D.:
  Curriculumnet: Weakly supervised learning from large-scale web images. In:
  Proceedings of the European conference on computer vision (ECCV). pp.
  135--150 (2018)

\bibitem{harel2023learning}
Harel, R., Pinter, Y., Oren, G.: Learning to parallelize in a shared-memory
  environment with transformers. In: Proceedings of the 28th ACM SIGPLAN Annual
  Symposium on Principles and Practice of Parallel Programming. pp. 450--452
  (2023)

\bibitem{harel2020source}
Harel, R., Mosseri, I., Levin, H., Alon, L.o., Rusanovsky, M., Oren, G.:
  Source-to-source parallelization compilers for scientific shared-memory
  multi-core and accelerated multiprocessing: analysis, pitfalls, enhancement
  and potential. International Journal of Parallel Programming  \textbf{48}(1),
   1--31 (2020)

\bibitem{hasabnis2021controlflag}
Hasabnis, N., Gottschlich, J.: Controlflag: A self-supervised idiosyncratic
  pattern detection system for software control structures. In: Proceedings of
  the 5th ACM SIGPLAN International Symposium on Machine Programming. p.
  32–42. MAPS 2021, Association for Computing Machinery, New York, NY, USA
  (2021). \doi{10.1145/3460945.3464954},
  \url{https://doi.org/10.1145/3460945.3464954}

\bibitem{henke2022semantic}
Henke, J., Ramakrishnan, G., Wang, Z., Albarghouth, A., Jha, S., Reps, T.:
  Semantic robustness of models of source code. In: 2022 IEEE International
  Conference on Software Analysis, Evolution and Reengineering (SANER). pp.
  526--537. IEEE (2022)

\bibitem{husain2019codesearchnet}
Husain, H., Wu, H.H., Gazit, T., Allamanis, M., Brockschmidt, M.: Codesearchnet
  challenge: Evaluating the state of semantic code search. arXiv preprint
  arXiv:1909.09436  (2019)

\bibitem{intel-cloud}
Intel: {Intel Developer Cloud}.
  \url{https://www.intel.com/content/www/us/en/developer/tools/devcloud/overview.html}
  (2023), [Online]

\bibitem{mosseri2020compar}
Mosseri, I., Alon, L., Harel, R., Oren, G.: Compar: Optimized multi-compiler
  for automatic openmp {S2S} parallelization. In: Milfeld, K.F., de~Supinski,
  B.R., Koesterke, L., Klinkenberg, J. (eds.) OpenMP: Portable Multi-Level
  Parallelism on Modern Systems - 16th International Workshop on OpenMP,
  {IWOMP} 2020, Austin, TX, USA, September 22-24, 2020, Proceedings. Lecture
  Notes in Computer Science, vol. 12295, pp. 247--262. Springer (2020).
  \doi{10.1007/978-3-030-58144-2\_16},
  \url{https://doi.org/10.1007/978-3-030-58144-2\_16}

\bibitem{neamtiu2005understanding}
Neamtiu, I., Foster, J.S., Hicks, M.: Understanding source code evolution using
  abstract syntax tree matching. ACM SIGSOFT Software Engineering Notes
  \textbf{30}(4), ~1--5 (2005)

\bibitem{niu2023empirical}
Niu, C., Li, C., Ng, V., Chen, D., Ge, J., Luo, B.: An empirical comparison of
  pre-trained models of source code. arXiv preprint arXiv:2302.04026  (2023)

\bibitem{niu2022spt}
Niu, C., Li, C., Ng, V., Ge, J., Huang, L., Luo, B.: {SPT-Code}:
  Sequence-to-sequence pre-training for learning the representation of source
  code. arXiv preprint arXiv:2201.01549  (2022)

\bibitem{parr2013definitive}
Parr, T.: The definitive ANTLR 4 reference. Pragmatic Bookshelf (2013)

\bibitem{platanios2019competence}
Platanios, E.A., Stretcu, O., Neubig, G., Poczos, B., Mitchell, T.M.:
  Competence-based curriculum learning for neural machine translation. arXiv
  preprint arXiv:1903.09848  (2019)

\bibitem{prema2017identifying}
Prema, S., Jehadeesan, R., Panigrahi, B.: Identifying pitfalls in automatic
  parallelization of nas parallel benchmarks. In: Parallel Computing
  Technologies (PARCOMPTECH), 2017 National Conference on. pp.~1--6. IEEE
  (2017)

\bibitem{prema2019study}
Prema, S., Nasre, R., Jehadeesan, R., Panigrahi, B.: A study on popular
  auto-parallelization frameworks. Concurrency and Computation: Practice and
  Experience  \textbf{31}(17),  e5168 (2019)

\bibitem{quiring2019misleading}
Quiring, E., Maier, A., Rieck, K., et~al.: Misleading authorship attribution of
  source code using adversarial learning. In: USENIX Security Symposium. pp.
  479--496 (2019)

\bibitem{rebuffi2021data}
Rebuffi, S.A., Gowal, S., Calian, D.A., Stimberg, F., Wiles, O., Mann, T.A.:
  Data augmentation can improve robustness. Advances in Neural Information
  Processing Systems  \textbf{34},  29935--29948 (2021)

\bibitem{vaswani2017attention}
Vaswani, A., Shazeer, N., Parmar, N., Uszkoreit, J., Jones, L., Gomez, A.N.,
  Kaiser, {\L}., Polosukhin, I.: Attention is all you need. Advances in neural
  information processing systems  \textbf{30} (2017)

\bibitem{DBLP:journals/corr/VaswaniSPUJGKP17}
Vaswani, A., Shazeer, N., Parmar, N., Uszkoreit, J., Jones, L., Gomez, A.N.,
  Kaiser, L., Polosukhin, I.: Attention is all you need. CoRR
  \textbf{abs/1706.03762} (2017), \url{http://arxiv.org/abs/1706.03762}

\bibitem{wang2022bridging}
Wang, D., Jia, Z., Li, S., Yu, Y., Xiong, Y., Dong, W., Liao, X.: Bridging
  pre-trained models and downstream tasks for source code understanding. In:
  Proceedings of the 44th International Conference on Software Engineering. pp.
  287--298 (2022)

\bibitem{DBLP:deepscc}
Yang, G., Zhou, Y., Yu, C., Chen, X.: Deepscc: Source code classification based
  on fine-tuned roberta. CoRR  \textbf{abs/2110.00914} (2021),
  \url{https://arxiv.org/abs/2110.00914}

\end{thebibliography}

\end{document}